\documentclass[
aps,
prb,
twocolumn,
]{revtex4-1}
\usepackage{amsmath}
\usepackage{amssymb}
\usepackage{accents}
\usepackage{mathrsfs}
\usepackage{color}
\usepackage{graphicx}
\definecolor{dgreen}{RGB}{00, 140, 00}
\usepackage{ulem}

\newcommand{\bs}[1]{\boldsymbol{#1}}
\newcommand{\eqdef}{\overset{\mathrm{def}}{=\joinrel=}}

\begin{document}

\title{Theory of tunneling spectroscopy of normal
metal/ferromagnet/spin-triplet superconductor junctions}

\author{L.A.B. Olde Olthof$^1$} 
\author{S.-I. Suzuki$^2$} 
\author{A.A. Golubov$^{1,3}$} 
\author{M. Kunieda$^4$} 
\author{S. Yonezawa$^4$} 
\author{Y. Maeno$^4$}
\author{Y. Tanaka$^2$\vspace{3mm}}

\affiliation{$^1$MESA$^+$ Institute for Nanotechnology, University of Twente, The Netherlands}
\affiliation{$^2$Department of Applied Physics, Nagoya University, Nagoya, 464-8603, Japan}
\affiliation{$^3$Moscow Institute of Physics and Technology, Dolgoprudny, Moscow Region, 141700, Russia}	
\affiliation{$^4$Department of Physics, Graduate School of Science, Kyoto University, Kyoto 606-8502, Japan}

\begin{abstract} 
We study the tunneling conductance of a ballistic normal metal /
ferromagnet / spin-triplet superconductor junction using the extended
Blonder-Tinkham-Klapwijk formalism as a model for a $c$-axis oriented Au /
SrRuO$_{3}$ / Sr$_{2}$RuO$_{4}$ junction.  We compare chiral $p$-wave (CPW)
and helical $p$-wave (HPW) pair potentials, combined with ferromagnet
magnetization directions parallel and perpendicular to the interface.  For
fixed $\theta_{M}$, where $\theta_{M}$ is a direction of magnetization in
the ferromagnet measured from the $c$-axis, the tunneling conductance of CPW and HPW clearly
show different voltage dependencies.  It is found that the cases where the
$d$-vector is perpendicular to the magnetization direction (CPW with 
$\theta_{M} = \pi/2$ and HPW with $\theta_{M} = 0$) are identical.  The
obtained results serve as a guide to determine the pairing symmetry of
the spin-triplet superconductor Sr$_{2}$RuO$_{4}$. 
\end{abstract}

\maketitle
\section{Introduction}
Nowadays, Sr$_{2}$RuO$_{4}$ is known as an
unconventional  superconductor with the transition temperature $T_{c} \sim
1.5$~K. \cite{Maeno94} The fact that the Knight shift does not change
across $T_c$ is consistent with spin-triplet pairing.\cite{Ishida, Maeno98,
Murakawa, Maeno2, Maeno2012} Various theoretical studies have discussed the
microscopic mechanism of spin-triplet pairings in this material. \cite{Rice,
Miyake, Ogata, Nomura00, Nomura02a, Nomura02b, Arita, Nomura05, Nomura08,
Yanase, Raghu, Kohmoto, Kuroki, Takimoto, Onari}  The existence of a zero
bias conductance peak in several tunneling experiments \cite{Laube, Mao}
indicates the realization of unconventional superconductivity.  \cite{TK95,
Kashiwaya00, YTK97}  In particular, the broad zero bias conductance peak
observed in tunneling spectroscopy suggests the realization of a surface
Andreev bound states (SABS) with linear dispersion \cite{Kashiwaya11,
YTK97, YTK98, Honerkamp}.  This is in contrast with high $T_c$ cuprate
superconductors, where a sharp zero bias conductance peak is observed
\cite{Kashiwaya00, TK95, Experiment1, Experiment2, Experiment3,
Experiment4, Experiment5, Experiment6} due to flat band zero energy
states.\cite{TK95, Hu, STYY11} When spin-triplet pairing is realized, we
can expect exotic phenomena, such as the so called anomalous proximity
effect  in diffusive normal metal / spin-triplet superconductor junctions.
\cite{Proximityp, Proximityp2, Proximityp3, Meissner3, Asano2011}

The presence or absence of time reversal symmetry (TRS) in
Sr$_{2}$RuO$_{4}$ is an important issue. Among two-dimensional spin-triplet
$p$-wave pairings, chiral and helical $p$-wave pairing seem promising in
the absence and presence of TRS, respectively.\cite{Qi11} Broken TRS was
observed in $\mu$SR and Kerr-rotation experiments as a result of a
spontaneous internal
magnetic field below $T_c$,\cite{Luke,Xia,Matsumoto99,Kerr} which supports
chiral $p$-wave pairing.  However, the internal magnetic field has not been
detected in scanning SQUID experiments,\cite{Kirtley,Hicks} which suggests
realization of helical $p$-wave symmetry.  Although there are several
possible explanations for the absence of broken TRS in Sr$_{2}$RuO$_{4}$,
\cite{Kallin1, Kallin2, Kallin3, Kallin4, Raghu,Sigrist201,Tada,Suzuki}
the pairing symmetry remains
a point of discussion.  One of the main differences between these two
pairing symmetries is the direction of $d$-vector.  

A constructive way to distinguish between them is to study the charge
transport in ferromagnet / spin-triplet superconductor
junctions.\cite{Hirai1, Hirai2, Yoshida, Li, Cheng2}
Naively speaking, the direction of the magnetization axis with respect to
the $d$-vector (parallel or perpendicular) influences the charge transport.
Recently, a Au / SrRuO$_{3}$ / Sr$_{2}$RuO$_{4}$ junction oriented along
the $c$-axis has been fabricated by means of epitaxial growth.\cite{Anwar2}
Since SrRuO$_{3}$ and Sr$_{2}$RuO$_{4}$ have similar $a$-axis lattice
constants, as well as similar atomic arrangements,
a smooth interface between them can be expected, which turns
this system into a nice playground for clarifying the direction of the
$d$-vector.  Because the SABS is absent in this direction, we can directly
compare the effect of the magnetization direction relative to the
$d$-vector.  To interpret the experimental results, a theoretical model is
required in which we calculate the tunneling conductance along the
$c$-axis, based on a minimal model which takes the quasi-two dimensional
nature of Sr$_{2}$RuO$_{4}$ into account. 

In this paper, we investigate normal metal (N) / ferromagnet (F) / spin-triplet superconductor (S) junctions with $s$-wave, chiral and helical $p$-wave pairing symmetries by changing the properties of the ferromagnet, $e.g.$, 
thickness, magnetization strength and direction.
The anisotropic Fermi surface of Sr$_2$RuO$_4$ and realistic effective masses are also included,
since the Fermi-momentum mismatch changes the transparency and the resulting conductance. 
Finally, an external magnetic field is taken into account through the Doppler shift.



\section{Formulation}
\subsection{Model and Hamiltonian}
\begin{figure}[b]
	\centering
		\includegraphics[scale=2]{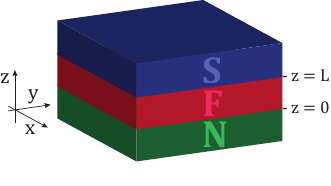}
		\vspace{-.5cm}
		\caption{Schematic of the three-dimensional normal
			metal(N)/ferromagnet(F)/superconductor(S) junction.
			We assume the structure extends to infinity in all directions.}
		\label{fig:NFSschematic}
\end{figure}

We consider a three-dimensional N/F/S junction, as shown in Fig.~\ref{fig:NFSschematic}. 
We assume the junction interfaces to be perpendicular to the $z$-axis
and located at $z=0$ and $z=L$. 
The F has a thickness $L$ and a magnetization $\boldsymbol{M}(z)$.
The N and S are considered to be semi-infinite.
Superconducting junctions are described by the Bogoliubov-de Gennes (BdG)
Hamiltonian
\begin{align}
	\mathcal{\check{H}}(\boldsymbol{r}) 
	&=
	\left[
		\begin{array}{cc}
			 \hat{h     }  ( \boldsymbol{r},H)    ~ &
			 \hat{\Delta}  ( \boldsymbol{r}) ~ \\[2mm]
			-\hat{\Delta}^*( \boldsymbol{r}) ~ &
			-\hat{h     }^*( \boldsymbol{r},H)    ~ \\
	\end{array}\right], 
\end{align}
where the basis is taken as $ \Psi(\boldsymbol{r}) = 
	[\,
   \psi_{\uparrow}           (\boldsymbol{r}) \hspace{2mm} 
	 \psi_{\downarrow}         (\boldsymbol{r}) \hspace{2mm} 
	 \psi_{\uparrow}  ^\dagger (\boldsymbol{r}) \hspace{2mm} 
   \psi_{\downarrow}^\dagger (\boldsymbol{r}) \,]^T, $
where $T$ is the transpose, the symbol $\hat{\cdot}$ ($\check{\cdot}$)
represents a $2 \times 2$ ($4 \times 4$) matrix in the spin (spin-Nambu)
space, $H$ is an externally applied magnetic field in the $x$-direction.
Since the system has translational symmetry in the $x$- and
$y$-direction, the momenta $k_x$ and $k_y$ are well-defined quantum
numbers.  Therefore, the wave function can be expressed in the Fourier
components as
\begin{align}
  &\Psi(\boldsymbol{r})
  =
  \sum_{\boldsymbol{k}_\parallel}
  \Psi_{\boldsymbol{k}_\parallel}(z)
  \frac{  e^{i (k_x x + k_y y) } }{\sqrt{L_x L_y}}, \label{fourier} \\
  &\Psi_{\boldsymbol{k}_\parallel}(z)
  =
	[~
   \psi_{  \uparrow, { \boldsymbol{k}_\parallel}}        ~ ~
	 \psi_{\downarrow, { \boldsymbol{k}_\parallel}}        ~ ~
	 \psi_{  \uparrow, {-\boldsymbol{k}_\parallel}}^\dagger~ ~
   \psi_{\downarrow, {-\boldsymbol{k}_\parallel}}^\dagger~ 
  ]^T,
\end{align}
where $\boldsymbol{k}_\parallel = (k_x, k_y, 0)$. 
In Eq.~(\ref{fourier}), we assume periodic boundary conditions in order to
accommodate the infinite dimensions in the $x$- and $y$-direction.
The lateral dimensions $L_x$ and $L_y$ are normalization factors
and do not affect the conductance spectrum. The Hamiltonian becomes
\begin{align}
	\mathcal{\check{H}}_{\bs{k}_\parallel}(z,H) 
	&=
	\left[
		\begin{array}{cc}
			 \hat{h     }  _{ \boldsymbol{k}_\parallel}(z,H)    ~ &
			 \hat{\Delta}  _{ \boldsymbol{k}_\parallel}(z) ~ \\[2mm]
			-\hat{\Delta}^*_{-\boldsymbol{k}_\parallel}(z) ~ &
			-\hat{h     }^*_{-\boldsymbol{k}_\parallel}(z,H)    ~ \\
	\end{array}\right]. 
\end{align}
The single-particle Hamiltonian $\hat{h}_{\boldsymbol{k}_\parallel}$ is given by 
\begin{align}
  &\hat{h}_{\boldsymbol{k}_\parallel}(z,H)
  = \xi_{\bs{k}_\parallel}(z,H) +\boldsymbol{M}(z) \cdot \boldsymbol{\hat{\sigma}} 
    + \hat{F}_{\bs{k}_\parallel}(z), \\[1mm]
 & \xi_{\bs{k}_\parallel}(z,H) 
 = - \frac{\hbar^2}{2m_z}\frac{\partial^2}{\partial z^2} - \mu'
 - \Delta_0\frac{H}{H_c} \frac{k_\parallel}{k_F}\sin\varphi, \label{eq:kinetic}  \\[1mm]
 &\mu'
 = \mu
 - \frac{\hbar^2}{2}
 \left[ \frac{k_x^2}{m_x} + \frac{k_y^2}{m_y} \right],
\end{align}
where $\xi_{\bs{k}}$ is the kinetic energy in the presence of an external magnetic
field in the $x$-direction and $\mu$ is the chemical potential, which we
assume to be constant across the junction. A full derivation of
Eq.~(\ref{eq:kinetic}) is given in Appendix~\ref{app:doppler}.  The
matrices $ \hat{\sigma}_{j}$ $(j\in \{x,y,z\})$ and $\hat{\sigma}_0$ are
the Pauli matrices and the identity matrix in spin space, 
$\hat{\boldsymbol{\sigma}} =  
  \hat{\sigma}_x \boldsymbol{    {e}}_x
+ \hat{\sigma}_y \boldsymbol{    {e}}_y
+ \hat{\sigma}_z \boldsymbol{    {e}}_z $ 
with $\boldsymbol{e}_{j}$ being the unit vectors in the $j$-direction.
We can modify the shape of Fermi surfaces by tuning the effective masses
$\boldsymbol{m} = (m_x, m_y, m_z)$ in each region. In this paper, we
parametrize $\boldsymbol{m}$ as 
\begin{align}
  \boldsymbol{m}(z) 
  = \left\{ \begin{array}{clc}
      (\, m_N         ,~ m_N        ,~ m_N \,)    & \hspace{5mm}\text{for}\hspace{1mm} & z \le 0,\\[1mm]
      (\, m_F         ,~ m_F        ,~ m_F \,)    & \hspace{5mm}\text{for}\hspace{1mm} & 0<z<L,\\[1mm]
      (\, m_\parallel ,~ m_\parallel,~ m_\perp\,) & \hspace{5mm}\text{for}\hspace{1mm} & z \ge L.\\
    \end{array} \right. 
\label{a}
\end{align}
The magnetization is described as \cite{Cheng2}
\begin{align}
  &\boldsymbol{M}(z)
  = M_0(\sin \theta_M \boldsymbol{e}_x +\cos \theta_M \boldsymbol{e}_z)
    \Theta(z) \Theta(L-z), 
\end{align}
where $\Theta(z)$ is the Heaviside step function. 
In this paper, we ignore the re-orientation of the
$d$-vector by the magnetization in F\cite{self1,self2,self3,self4} for simplicity.
The effects of the interfaces 
are described by $\hat{F}_{\boldsymbol{k}_\parallel}(z)$ as 
\cite{Samokhin}
\begin{align}
  &\hat{F}_{\boldsymbol{k}_\parallel}(z)
  = \delta(z) \hat{F}_1 +\delta(z-L) \hat{F}_2, \\[2mm]
  &\hat{F}_{1,\boldsymbol{k}_\parallel} = F_{1}\hat{\sigma}_0 , \\[1mm]
  &\hat{F}_{2,\boldsymbol{k}_\parallel} = F_{\mathrm{SO}} \boldsymbol{e}_z \cdot 
        ( \hat{\boldsymbol{\sigma}} \times \boldsymbol{k} ), 
\label{aa}
\end{align}
where $F_{1}$ and $F_{\mathrm{SO}}$ represent the strengths
of the barrier potential at $z=0$ and the spin-orbit coupling (SOC) at $z=L$, respectively.  The SOC term reduces to 
\begin{align}
  \boldsymbol{e}_z \cdot ( \hat{\boldsymbol{\sigma}} \times \boldsymbol{k} ) 
  &= \hat{\sigma}_x k_y - \hat{\sigma}_y k_x \\[1mm]
  &= i k_\parallel 
  \left[ \begin{array}{cc}
    0            & e^{-i \varphi} \\
    -e^{+i \varphi} & 0           \\
  \end{array} \right], 
\end{align}
where $k_x = k_\parallel \cos \varphi$ and $k_x = k_\parallel \sin \varphi$
with $k_\parallel = ( k_x^2 + k_y^2 )^{1/2}$. 
The pair potential is described by 
\begin{align}
  &\hat{\Delta}_{\boldsymbol{k}_\parallel}(z) 
  = \underline{\hat{\Delta}}_{\boldsymbol{k}_\parallel}(z) \Theta ( z-L ). 
\end{align}
The momentum dependence of
the pair potentials for $s$\,-wave (SW), chiral $p$\,-wave (CPW), and
helical $p$\,-wave (HPW) superconductors are written as
\begin{align}
  &\underline{\hat{\Delta}}_{\boldsymbol{k}_\parallel} (z)
  = \left\{ \begin{array}{ll}
      \Delta_0 i \hat{\sigma}_y & \text{for SW, } \\[2mm]
      \Delta_0
      [ \bar{k}_x + i \chi \bar{k}_y ] 
      \hat{\sigma}_x & \text{for CPW, } \\[2mm]
      \Delta_0
			[ \bar{k}_x \hat{\sigma}_0
			+ i \bar{k}_y \hat{\sigma}_z] & 
      \text{for HPW, } \\
    \end{array} \right. \label{eq:pp}
\end{align} 
where $\Delta_0$ is a constant which characterizes the amplitude of the pair
potential, $\chi$ is so-called the chirality (which can be $\pm 1$),
$\bar{k}_{x} = k_x/k_{s\parallel}$ with $k_{s\parallel} =
\sqrt{2m_{\parallel} \mu'}/\hbar$ being the Fermi wavenumber in $k_x$-$k_y$
plane for S. 
The assumption that $\Delta_0$ is constant implies that we do not take
the inverse proximity effect (from F into S) into account, which is a common assumption.\cite{btk}

\subsection{Wave functions}
The wave function is obtained by solving the Hamiltonian at an energy $E$ in each region. 
Throughout this paper, we assume $E \sim \Delta_0 \ll \mu$. 
The wave function for $z \le 0$ is given by 
\begin{align}
  \Psi_{k_\parallel}(z) 
  =
  \check{K}^+_N \vec{i} + \check{K}_N^- \vec{r}, 
  \label{eq:waveN}
\end{align}
where $\check{K}^\pm_N = e^{\pm i \check{\tau}_z k_N z}$ with $k_N = \sqrt{2m_N \mu'}/\hbar$ 
and $\check{\tau}_z = \mathrm{diag} [ \hat{\sigma}_0, -\hat{\sigma}_0 ] $ being the third Pauli matrix in Nambu space. The vector $\vec{i}$ represents the
wave function amplitude of the incident particles  which is given by
\begin{align}
  \vec{i} = 
  \left\{ \begin{array}{cl}
   \left[~1~0~0~0~\right]^T ~ & ~\text{for an  up-spin electron},  \\[2mm]
   \left[~0~1~0~0~\right]^T ~ & ~\text{for a down-spin electron}.  \\
  \end{array} \right. 
\end{align}
The vector $\vec{r}$ describes the
wave function amplitude of the reflected particles as
\begin{align}
  \vec{r} = 
	[~
    r^p_{  \uparrow} ~ ~
    r^p_{\downarrow} ~ ~
    r^h_{  \uparrow} ~ ~
	r^h_{\downarrow} ~  
  ]^T, \label{eq:rcoef}
\end{align}
where $r^p_\alpha$ and $r^h_\alpha$, $\alpha \in \{ \uparrow, \downarrow \}$ are the normal
and Andreev reflection coefficients, respectively. The wave function for $0<z<L$ is given
by 
\begin{align}
	\Psi_{\bs{k}_\parallel}(z) = 
	 \check{A} \check{K}^+ _F  \vec{f}_P
	+ \check{A}\check{K}^-_F \vec{f}_N, 
	\label{eq:waveF}
\end{align}
where $\check{K}_F^\pm = \mathrm{diag}[\,
e^{\pm ik_F^+z}, ~
e^{\pm ik_F^-z}, ~
e^{\pm ik_F^+z}, ~
e^{\pm ik_F^-z}\,]$ with $k_F^\pm = \sqrt{2m_F(\mu' \mp M_0)}/\hbar$. 
The matrix $\check{A} = \mathrm{diag}[ \hat{A}, \hat{A}]$ characterizes the 
spin structure of the F, where $\hat{A}$ is given by \cite{Cheng2}
\begin{align}
  \hat{A}= 
	\left[ \begin{array}{rr}
    ~ \cos ( {\theta_M}/{2} ) & 
    ~-\sin ( {\theta_M}/{2} ) ~\\[2mm]
    ~ \sin ( {\theta_M}/{2} ) & 
    ~ \cos ( {\theta_M}/{2} ) ~ \\
	\end{array} \right]
\end{align}
The vectors $\vec{f}_{P(N)}$ describe the wave function amplitudes of particles propagating in the positive (negative) $z$-direction. They are
defined as 
\begin{align}
  \vec{f}_P =& \,
	[~
    f^p_{  \uparrow,P} ~ ~
    f^p_{\downarrow,P} ~ ~
    f^h_{  \uparrow,P} ~ ~
	  f^h_{\downarrow,P} ~  
  ]^T, \\[1mm]
  \vec{f}_N =& \,
	[~
    f^p_{  \uparrow,N} ~ ~
    f^p_{\downarrow,N} ~ ~
    f^h_{  \uparrow,N} ~ ~
	  f^h_{\downarrow,N} ~  
  ]^T. 
\end{align}
The wave function for $z \ge L$ is given by 
\begin{align}
\Psi_{k_\parallel}(z) = 
\check{U} \check{K}_S \vec{t}, 
\label{eq:waveS}
\end{align}	
where
$\check{K}_S = \mathrm{diag}[~
e^{+i k_S z}, \, ~
e^{+i k_S z}, \, ~
e^{-i k_S z}, \, ~
e^{-i k_S z}~]$ with $k_S = \sqrt{2m_\perp \mu'}/\hbar$. 
The vector $\vec{t}$ describes the wave function amplitudes
of the transmitted particles as
\begin{align}
  \vec{t} =& \,
	[~
    t^{p}_{  \uparrow} ~ ~
    t^{p}_{\downarrow} ~ ~
    t^{h}_{  \uparrow} ~ ~
	  t^{h}_{\downarrow} ~  
  ]^T. 
\end{align}
The matrix $\check{U}$ describes the amplitude of the wave
function in the superconductor as 
\begin{align}
  &\check{U}
  = 
	\left[ \begin{array}{ll}
    u_{\boldsymbol{k}_\parallel} \hat{\sigma}_0 & ~ 
    v_{\boldsymbol{k}_\parallel} \, \hat{D}_{\boldsymbol{k}_\parallel}         \\[2mm]
	  v_{\boldsymbol{k}_\parallel} \, \hat{D}_{\boldsymbol{k}_\parallel}^\dagger & ~ 
    u_{\boldsymbol{k}_\parallel} \hat{\sigma}_0 \\
  \end{array} \right], \\[2mm]
  & \hat{D}_{\boldsymbol{k}_\parallel}
  = \hat{\underline{\Delta}}_{\boldsymbol{k}_\parallel}
   /\Delta_0, 
\end{align}
with
\begin{align}
  & u_{\boldsymbol{k}_\parallel}
  = \frac{1}{\sqrt{2}} \, \sqrt{1 + \frac{\Omega_{\boldsymbol{k}_\parallel}}{E}}, \\
  & v_{\boldsymbol{k}_\parallel}
  = \frac{1}{\sqrt{2}} \, \sqrt{1 - \frac{\Omega_{\boldsymbol{k}_\parallel}}{E}}, \\[1mm]
  &\Omega_{\boldsymbol{k}_\parallel}
  = \sqrt{ E^2 - |d_{\boldsymbol{k}_\parallel}|^2}, 
\end{align}
where $d_{\boldsymbol{k}_\parallel}$ is obtained from the relation $
d_{\boldsymbol{k}_\parallel} \hat{\sigma}_0 = 
    \hat{\underline{\Delta}}_{\boldsymbol{k}_\parallel}   
    \hat{\underline{\Delta}}_{\boldsymbol{k}_\parallel}^\dagger $.

\subsection{Differential conductance}

\begin{figure*}[tb]
	\centering
	\includegraphics[width=\linewidth]{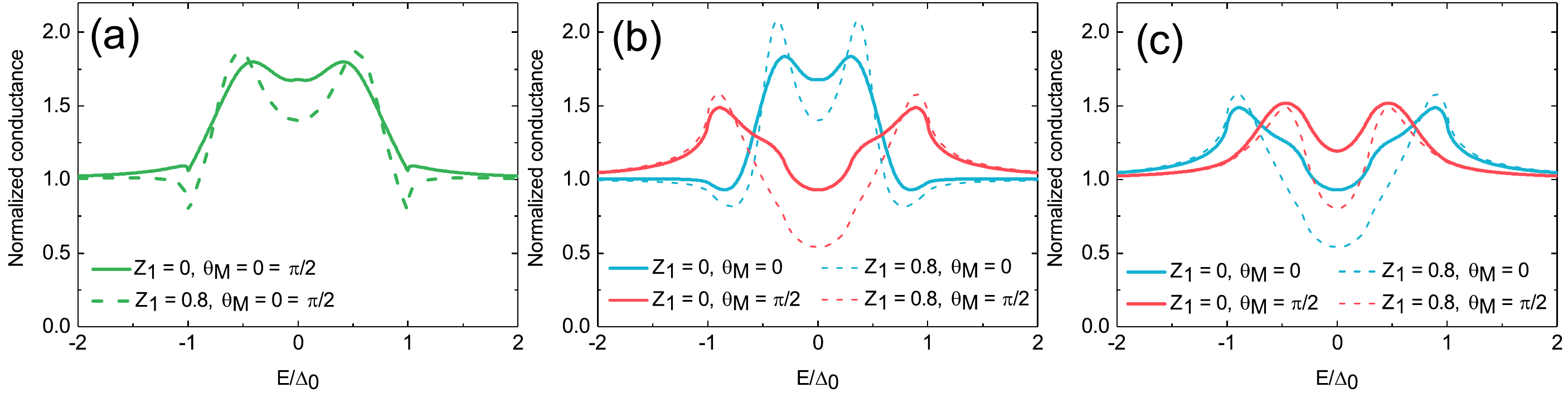}
	\caption{The dimensionless tunneling conductance using pair potentials
	{\bf (a)} SW, {\bf (b)} CPW and {\bf (c)} HPW.  Without barriers
	($Z_1=0$, solid graphs) and including a small barrier at the first
	interface ($Z_1=0.8$, dashed graphs).  The SW case is independent of the
	magnetization angle.  For CPW and HPW, the magnetization angle varies
	from $\theta_M=0$ (blue graphs) to $\theta_M=\pi/2$ (red graphs).  $X =
	0.6$, $k_FL = 11$.  }
	\label{fig:Z1dependence}
\end{figure*}
\begin{table*}[tb]
	\centering
	\includegraphics[scale=1.2]{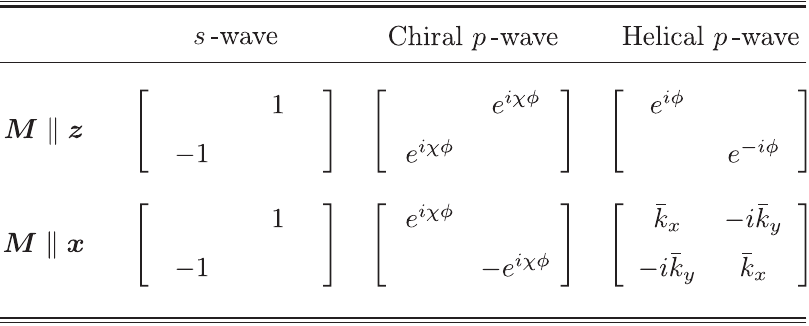}
	\caption{Matrix structure of the pair potential. The spin-quantisation
		axis is taken to be parallel to the magnetisation vector $\bs{M}$.
		The	first and second rows are for $\bs{M} \parallel \bs{z}$
		(i.e., $\theta_M
		= 0$) and for $\bs{M} \parallel \bs{x}$
		(i.e., $\theta_M = \pi /2$),
		respectively. From the table, we can see that the $4 \times 4$
		Hamiltonian can be reduced to two $2 \times 2$ Hamiltonian matrices
		except for the helical $p$\,-wave with $\theta_M = \pi /2$ case.
    The angle $\phi$ satisfies the relations: ${k}_{x} = k_\parallel \cos
		\phi$ and ${k}_{y} = k_\parallel \sin \phi$ with $k_{\parallel} =
		|\bs{k}_{\parallel}|$ being the momentum parallel to the interfaces. 
    The momentum is normalized: $\bar{k}_{x(y)} = k_{x(y)}/k_{\parallel}$. 
	  The factor $\chi$ is the chirality of a chiral $p$-wave superconductor. }
	\label{tablepairpotential}
\end{table*} 

All coefficients in Eq.~(\ref{eq:waveN}), (\ref{eq:waveF}) and
(\ref{eq:waveS}) can be determined by the four boundary conditions at $z=0$
and $z=L$. The first two boundary conditions are derived from continuity at
$z=0$. They are given by\cite{btk}
\begin{align}
  & \lim_{z  \uparrow 0} \Psi_{\boldsymbol{k}_\parallel} 
  = \lim_{z\downarrow 0} \Psi_{\boldsymbol{k}_\parallel}, 
  \label{eq:bc1} \\[1mm]
  & \lim_{z  \uparrow 0} \left[ 
      \frac{\partial           \Psi_{\boldsymbol{k}_\parallel}}{\partial z} 
      + \frac{2 m(z)}{\hbar^2} \check{F}_1 \Psi_{\boldsymbol{k}_\parallel}
    \right] 
  = \lim_{z\downarrow 0} \frac{\partial \Psi_{\boldsymbol{k}_\parallel}}{\partial z},  \label{eq:bc2}
\end{align}
where $\check{F}_1 = \mathrm{diag}[ \hat{F}_{1,\boldsymbol{k}_\parallel},
-\hat{F}_{1,-\boldsymbol{k}_\parallel}^*]$.  The other boundary conditions
are related to the interface at $z=L$ as follows
\begin{align}
  & \lim_{z  \uparrow L} \Psi_{\boldsymbol{k}_\parallel} 
  = \lim_{z\downarrow L} \Psi_{\boldsymbol{k}_\parallel},
  \label{eq:bc3} \\[1mm]
  & \lim_{z  \uparrow L} \left[ 
      \frac{\partial           \Psi_{\boldsymbol{k}_\parallel}}{\partial z} 
      + \frac{2 m(z)}{\hbar^2} \check{F}_2 \Psi_{\boldsymbol{k}_\parallel}
    \right] 
  = \lim_{z\downarrow L} \frac{\partial \Psi_{\boldsymbol{k}_\parallel}}{\partial z}, \label{eq:bc4}
\end{align}
where $\check{F}_2 = \mathrm{diag}[ \hat{F}_{2,\boldsymbol{k}_\parallel},
-\hat{F}_{2,-\boldsymbol{k}_\parallel}^*]$. 

In the supplemental material, we derived the expression for the current
through the N/F/S junction and found that it was the same as in the
original BTK theory.\cite{btk} Hence, we can use the same differential tunnelling conductance resulting from a spin-$\alpha$ incident particle,
which is given by
\begin{align}
  &{\sigma}(E) 
  = \sum_{\boldsymbol{k}_\parallel, \alpha} {}^{'}
    {\sigma}_{\alpha}(E, \boldsymbol{k}_\parallel), \label{sumsigma} \\
  &{\sigma}_{\alpha}(E, \boldsymbol{k}_\parallel)
  = 1 + |r^h_{  \uparrow}|^2 
      + |r^h_{\downarrow}|^2 
      - |r^p_{  \uparrow}|^2 
      - |r^p_{\downarrow}|^2, 
			\label{eq:cond}
\end{align}
where ${\sigma}_{\alpha}(E, \boldsymbol{k}_\parallel)$ is the
angle-resolved differential conductance for a spin-$\alpha$ incident
particle with $\alpha \in \{\uparrow, \, \downarrow \}$.\cite{notesum}
To model a cylindrical Fermi surface in a quasi-two-dimensional
material, we introduce a cutoff in the summation with respect to
$\boldsymbol{k}_\parallel$ as 
\begin{align}
  \sum_{\boldsymbol{k}_\parallel, \alpha} {}^{'} \cdots ~
  \eqdef ~
  \sum_{\boldsymbol{k}_\parallel, \alpha} \cdots
  \Theta \big( |\boldsymbol{k}_\parallel| - k_c \big)
  \label{eq:summation}
\end{align}
where $k_c = k_N \sin \theta_c$ and $\theta_c$ is the cutoff angle.

\begin{figure*}[tb]
	\centering
	\includegraphics[width=0.9\textwidth]{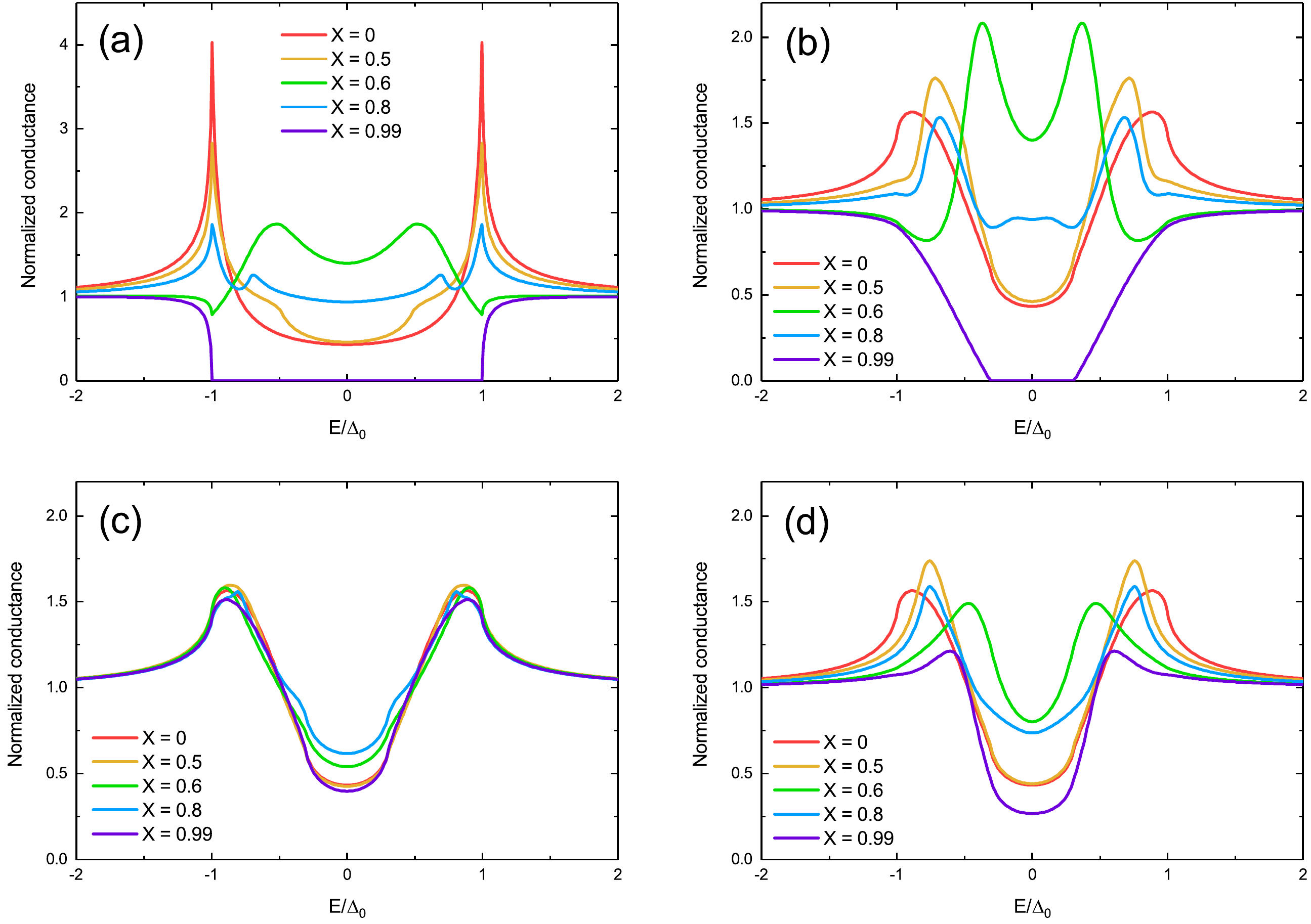}
	\caption{Dimensionless tunneling conductance using pair potentials
	{\bf (a)} SW, {\bf (b)} CPW, $\theta_M = 0$, {\bf (c)} HPW, $\theta_M =
	0$ which is identical to CPW, $\theta_M = \pi/2$ and {\bf (d)} HPW,
	$\theta_M = \pi/2$.  Magnetization strengths varying from $X=0$ (normal
	metal) to $X=0.99$ (fully polarized).  $Z_1 = 0.8$, $k_FL = 11$.  }
	\label{fig:magnetization}
\end{figure*} 

\section{Results}
The aim of this paper is to model the conductance of a
Au/SrRuO$_3$/Sr$_2$RuO$_4$ junction.  A realistic effective mass for
ferromagnet SrRuO$_3$ is $m_F = 7m_N$.\cite{SROmass} We approximate the
Sr$_2$RuO$_4$ $\gamma$-band by modelling the Fermi surface as an ellipsoid
($m_\parallel = 1.3$, $m_\bot = 16$) with its top and bottom cut off
($\theta_c = \pi/10$).  We will compare a N/F/S junction without barriers
to a N/F/S with a small tunnel barrier $F_1$
at the N/F interface.  Because of
epitaxial growth and minimal lattice mismatch, a smooth F/S interface is
expected and therefore, no barrier is introduced.
The spin-orbit coupling is set to zero in the main text of the main
text. Effects of $F_{SO}$ are discussed in Appendix~\ref{appendixrotate}.

\subsection{Direction of the magnetization}

We first show the differential conductances of a junction with a
spin-singlet $s$\,-wave superconductor in Fig.~\ref{fig:Z1dependence}(a), where results
with and without the interface barrier are indicated by solid and
dashed lines, respectively.
Throughout this paper, the differential
conductance is normalized by its value in the normal state
 (i.e., $\Delta_0 = 0$) and the energy is normalized by
 the maximum amplitude of the pair potential in the absence of an external magnetic field, $\Delta_0$. As shown
in Fig.~\ref{fig:Z1dependence}(a), the coherence peaks appear at an energy
lower than the gap amplitude ($E \approx 0.6 \Delta_0$)
which is a result of the ferromagnet with  finite thickness $L$. Comparing the solid and dashed lines, we see that the barrier
potential at the N/F interface sharpens the peaks around $E
\approx 0.5\Delta_0$ and the dips around $E \approx \Delta_0$
in the differential conductance.
In addition, the zero-energy dip becomes more prominent 
with increasing barrier. This is consistent with the well-known 
N/S junction.\cite{btk} In spin-singlet superconductors
the conductance does not depend on
the direction of the magnetization (i.e., $\theta_M$) because a singlet
Cooper pair does not have a finite total spin.
It should be noted that, throughout this paper,
	the pair potential is taken non-self-consistent (i.e., $\Delta_0$ is constant).
	The sharp peaks in the conductance would be broadened and lowered if we would
	include the self-consistency.\cite{halterman}

The differential conductance of the spin-triplet CPW and HPW
superconductors are shown
in Figs.~\ref{fig:Z1dependence}(b) and \ref{fig:Z1dependence}(c),
respectively. The blue and red lines represent the results
for $\theta_M = 0$ and $\pi/2$, respectively.

The cases with and without
N/F interface barrier are indicated by the solid and dashed lines,
respectively.
The results of the CPW, $\theta_M=0$ case are similar to the SW case;
 there are two peaks around
$E \approx 0.6 \Delta_0$ and a dip at zero-energy.  
The position of the peaks is determined by the F thickness ($L$)
and the magnitude of the magnetization ($X \equiv M/\mu$). 
In the CPW case, the Hamiltonian becomes equivalent to that
for the SW case, except for the amplitude of the pair potential.
Therefore, the corresponding results are qualitatively the same.

\begin{figure*}[tb]
	\centering
	\includegraphics[width=0.9\textwidth]{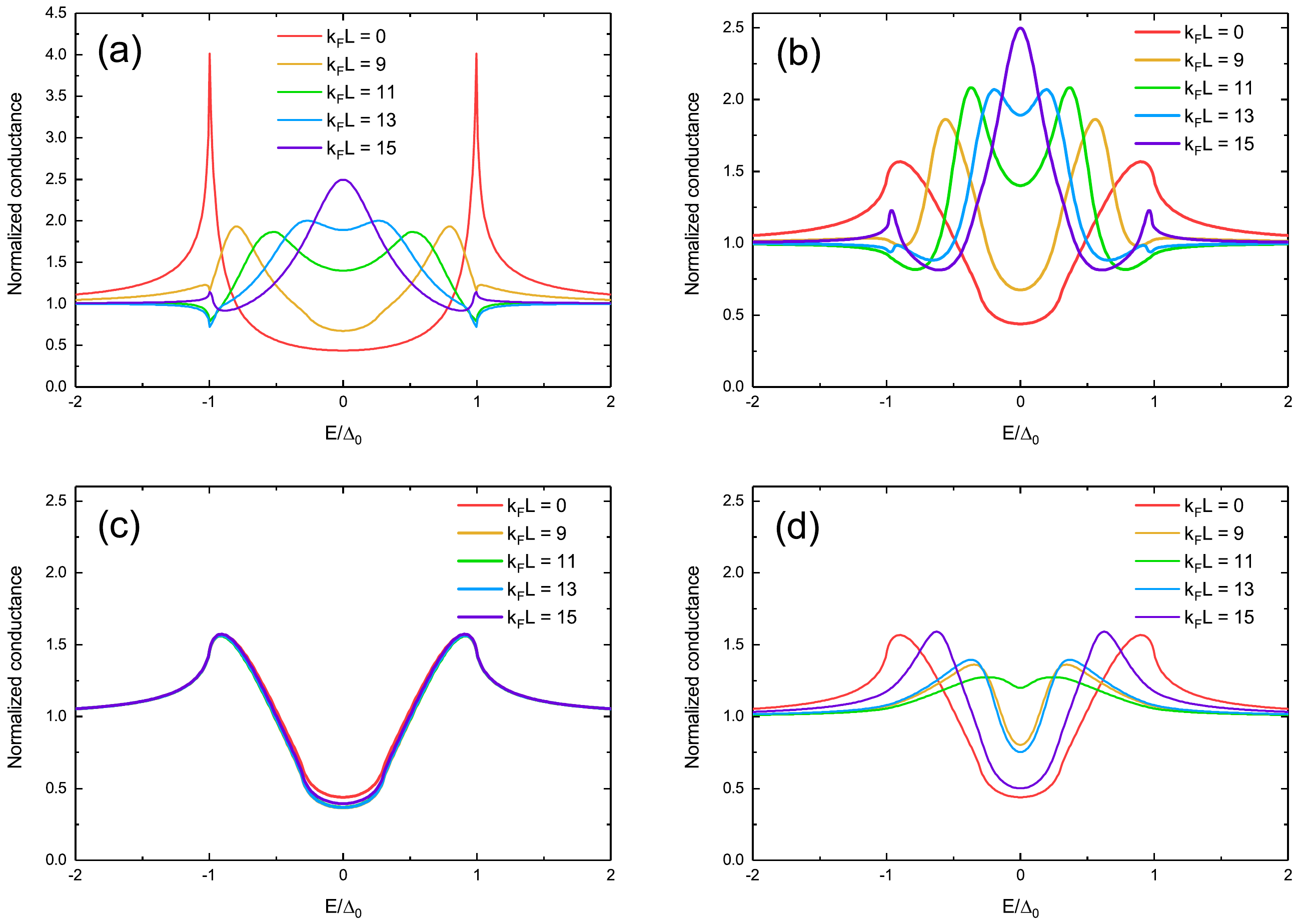}
	\caption{The dimensionless tunneling conductance using pair potentials
	{\bf (a)} SW, {\bf (b)} CPW, $\theta_M = 0$, {\bf (c)} HPW, $\theta_M =
	0$ which is identical to CPW, $\theta_M = \pi/2$ and {\bf (d)} HPW,
	$\theta_M = \pi/2$.  Without ferromagnet ($k_FL=0$) and for varying
	thicknesses $k_FL$ of the ferromagnet.  $Z_1 = 0.8$, $X=0.6$.  }
	\label{fig:length}
\end{figure*} 

In the present case, 
the experimentally observed zero-bias conductance peak
(ZBCP)\cite{Laube, Mao}
does not appear.
The Andreev bound states in CPW and HPW superconductors
are located in the $b$-$c$ and $c$-$a$ planes. The junction under consideration is, however, along the $c$-axis,
implying that these Andreev bound states cannot contribute to the
differential conductance.\cite{Note} 

Comparing the red line in Fig.~\ref{fig:Z1dependence}(b) to
 the blue line
in Fig.~\ref{fig:Z1dependence}(c), we find that the conductance spectra of
CPW with $\theta_M=\pi/2$ and HPW with $\theta_M=0$ are identical.
In both cases, the $d$-vector is perpendicular to the magnetization
($\vec{d} \ \bot \ \bs{M}$), i.e., the total spin of the
Cooper pairs is parallel to the magnetization.  

By analytically rotating the spin quantization axis, 
we reduce the matrix form of the pair potential matrix in the proper spin
axis in which the $z$-direction is parallel to the magnetization.  
By doing this, we demonstrate that the pair potentials in the
CPW, $\theta_M=\pi/2$ and HPW, $\theta_M=0$ cases are qualitatively the same, except for the spin-dependent chirality. A full
derivation is given in Appendix~\ref{appendixrotate}; the matrix structures
of the pair potential are summarized in table~\ref{tablepairpotential}. 
Hence, as far as there
is no perturbation which mixes the spins or depends on the chirality
(e.g., spin-active interface, spin-orbit coupling, or perturbation which
breaks translational symmetry in $x$ and/or $y$ direction such as walls
and impurities), it is impossible to distinguish between these two cases. 



\begin{figure*}[tb]
	\centering
	\includegraphics[width=0.78\textwidth]{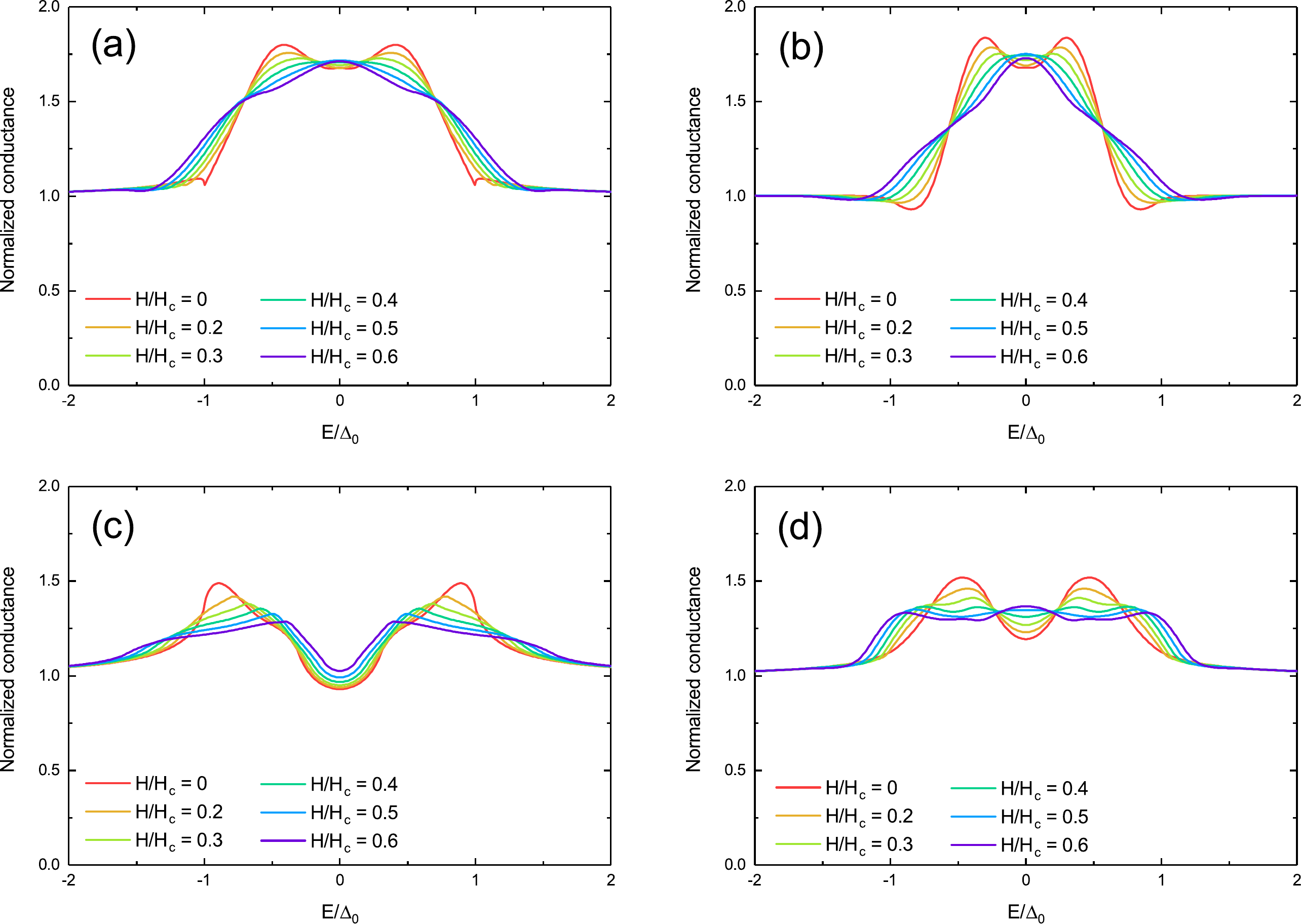}
	\caption{Effects of an external magnetic field on the dimensionless
	tunneling conductance in the \textit{absence} of the barrier potential. 
		The pair potential is assumed to be 
		{\bf (a)} SW, 
		{\bf (b)} CPW, $\theta_M = 0$, 
		{\bf (c)} HPW, $\theta_M = 0$, and 
		{\bf (d)} HPW, $\theta_M = \pi/2$. The results for the CPW with 
		$\theta_M = \pi/2$ are identical to the results in the panel (c). The
		parameters are set to $Z_1=0$, $X=0.6$, and $k_FL = 11$.  }
	\label{fig:magfield_nobar}
\end{figure*} 

\begin{figure*}[tb]
	\centering
	\includegraphics[width=0.78\textwidth]{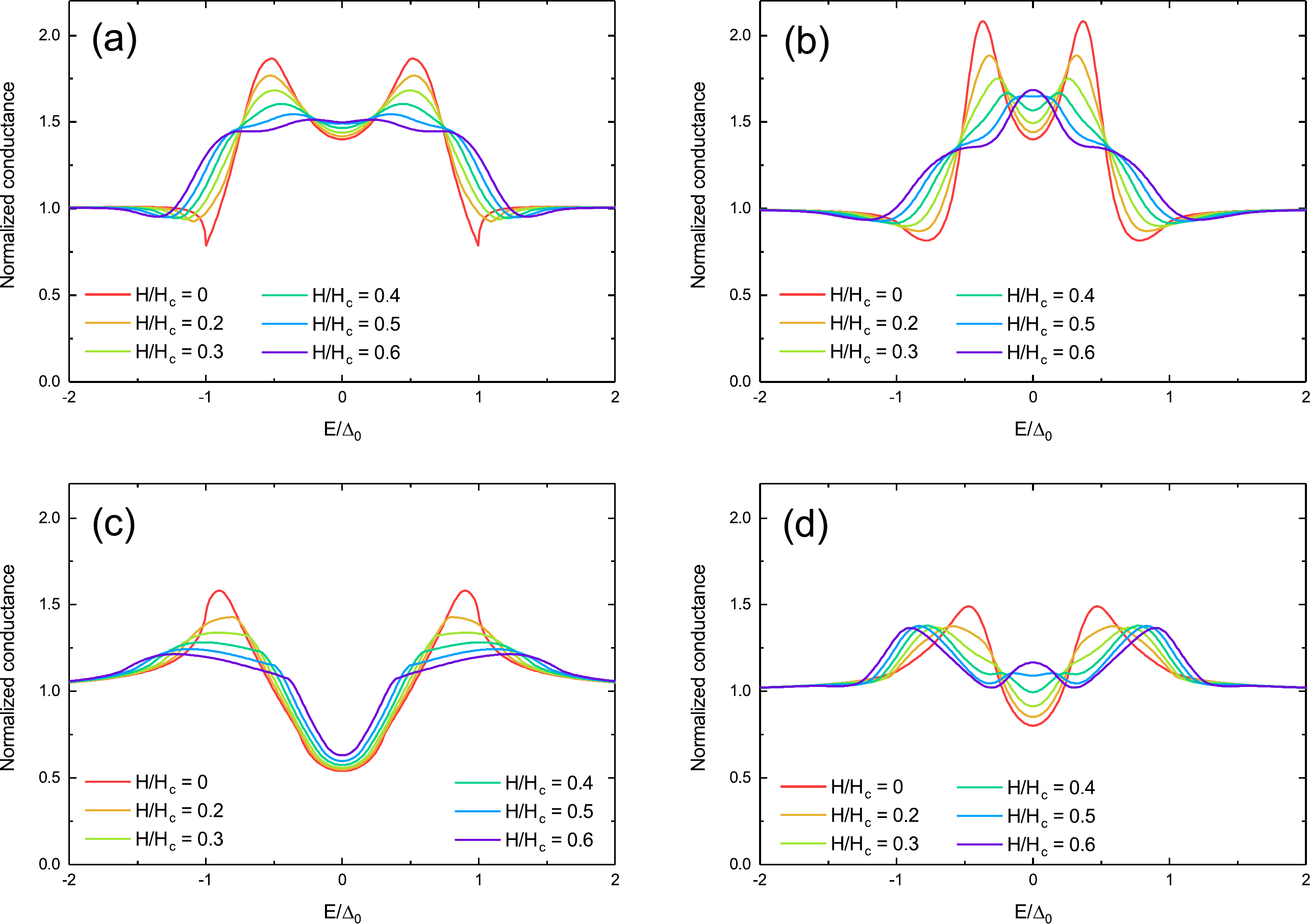}
	\caption{Effects of an external magnetic field on the dimensionless
	tunneling conductance in the \textit{presence} of the barrier potential $Z_1 =0.8$
	in the same manner as in Fig.~\ref{fig:magfield_nobar}. 
		}
	\label{fig:magfield_inclbar}
\end{figure*} 

\subsection{Amplitude of the magnetization}
The effects of the amplitude of the magnetization are shown in
Fig.~\ref{fig:magnetization}, where the pair potential and the direction of
the magnetization are set to 
(a)  SW with $\theta_M=0$, 
(b) CPW with $\theta_M=0$, 
(c) CPW with $\theta_M=\pi/2$, and
(d) HPW with $\theta_M=\pi/2$. 
We note that the result for the CPW with $\theta_M=\pi/2$ and that the HPW
with $\theta_M=0$ are identical to each other.
The barrier strength and the thickness
of the ferromagnet are set to $Z_1 = 0.8$ and $k_FL=11$, respectively. 

In the SW case in the absence of magnetization
($X=0$), we obtain the BTK-like U-shaped
spectrum\cite{btk} as shown in Fig.~\ref{fig:magnetization}(a).
Since the system is regarded as a N/N/S junction when $X=0$,
this result is well understood within the BTK theory. When the ferromagnet is fully
spin polarised ($X \approx 1$), the conductance becomes zero in the
energy range $|E| < \Delta_0$. 
Since there is no propagating channel in the S, a quasi particle
with energy $|E| < \Delta_0$ must be either normally or Andreev reflected at the F/S interface. In spin-singlet superconductors,
Andreev reflection is always accompanied
by a spin flip (e.g., an up-spin particle is reflected as a down-spin
hole). On the other hand, there is only one band in a fully-polarized
ferromagnet, which implies that Andreev reflection is prohibited. 
As a result, the conductance in the energy range $|E| < \Delta_0$ is always zero. 
For moderate
spin polarizations, the conductance spectra have complex structures that are sensitive to the amplitude of $\bs{M}$. 

The conductance spectrum in the CPW, $\theta_M=0$ case
[Fig.~\ref{fig:magnetization}(b)] is qualitatively the same as
the SW spectrum, because Cooper pairs
consist of quasi particles with opposite spin.
However, the CPW conductance changes more gradually as a
function of magnetization because the amplitude of the pair potential
changes depending on $k_z$. 
In the cases where $\vec{d} \ \bot \ \bs{M}$
[Fig.~\ref{fig:magnetization}(c)],
the conductance spectra do not depend on $\bs{M}$
qualitatively because the total spin of the Cooper 
pairs aligns with the magnetization.
This implies that the presence of the ferromagnet does not
affect the superconductivity and therefore, the conductance spectra
are insensitive to the magnetization.
Contrary to Figs.~\ref{fig:magnetization}(a) and
\ref{fig:magnetization}(b), the conductance in the HPW, $\theta_M=\pi/2$
 case [Fig.~\ref{fig:magnetization}(d)] remains finite even if $X \approx 1$.
In HPW superconductors, the $d$-vector lies in the $xy$-plain
in spin space. Therefore,
the $\bs{k}_\parallel$ dependent part of the Andreev reflection
is suppressed by the magnetization in the $x$-direction.

\subsection{Thickness of the ferromagnet}

In Fig.~\ref{fig:length}, the conductance spectra are plotted for
several thicknesses of the ferromagnetic layer $L$.
In the SW junction
[Fig.~\ref{fig:length}(a)], the conductance shows the BTK-like U-shaped spectrum\cite{btk}
as seen in Fig.~\ref{fig:magnetization}(a) with $X=0$.  The distance between the two peaks decreases with increasing thickness.
Simultaneously, the structures at $E=\Delta_0$ change from peaks to
dips. 
When $k_FL=15$, the two peaks merge into a ZBCP. 
We note that this peak is different from the well-known ZBCP in $d$-wave superconductors, which stems from the interference between
incident and reflected quasi particles at the interface.
On the other hand, the peak at the zero-energy in
Fig.~\ref{fig:magnetization}(a) is formed by an accidental constructive 
Fabry-Perot interference in the ferromagnet.\cite{interference}
Hence, this peak is not robustly resistant to impurities
and is therefore not related the topology in the superconductor.

Similar behaviour is seen in the spectrum of the CPW with $\theta_M=0$
case [Fig.~\ref{fig:length}(b)]. 
In HPW superconductors [Fig.~\ref{fig:length}(d)], 
the distance between the two peaks first reduces for $0 \leq k_F L \leq 11$, whereas it increases for $11 \leq k_F L \leq 15$. However, 
the constructive interference as seen in CPW superconductors never
occurs at the zero energy. This is a significant difference between 
CPW and HPW superconductors. 

When the $d$-vector is perpendicular to the magnetization
(i.e., $\vec{d} \ \bot \ \bs{M}$), the results are insensitive to the ferromagnet thickness,
as shown in Fig.~\ref{fig:length}(c). 
This can also be interpreted in
terms of the relation between the direction of $\bs{M}$ and the total
spin of Cooper pairs in the superconductor.

\subsection{External magnetic field}

The magnetic field dependence of the conductance in the
absence (presence) of a barrier at the N/F interface is shown in Fig.~\ref{fig:magfield_nobar}
(Fig.~\ref{fig:magfield_inclbar}), where the other parameters are set to
the same values used in Fig.~\ref{fig:magnetization}.  The pair potential
is assumed to be 
(a) SW, 
(b) CPW with $\theta_M = 0$, 
(c) CPW with $\theta_M = \pi/2$, and 
(d) HPW with $\theta_M = \pi/2$, where the results for the HPW with $\theta_M =
0$ are identical to the results in panel (c).
We show only the results for an external field $H \leq 0.6 H_c$,
since the effects of the
nucleation of vortices are not taken into account. 

In general, the Doppler shift causes peaks to
	split into two smaller peaks, which shift with $k_\parallel$,
	as follows from Eq.~(\ref{eqdoppler}).
Since pairing symmetries have different $k_\parallel$-dependencies,
the evolution of the peak shape is different in each case.
Both SW and CPW with $\theta_M=0$ 
[Figs.~\ref{fig:magfield_nobar}(a) and \ref{fig:magfield_nobar}(b)]
show a three dip structure that gradually transitions into a broad ZBCP.
For the CPW with $\theta_M=\pi/2$ and HPW with $\theta_M=0$ cases
[Fig.~\ref{fig:magfield_nobar}(c)], the coherence peaks are smeared out by
the magnetic field, although the central dip remains.
In the HPW with $\theta_M=\pi/2$ case [Fig.~\ref{fig:magfield_nobar}(d)],
the two peaks are split into four smaller peaks [$H/H_c=0.4$ in Fig.~\ref{fig:magfield_nobar}(d)]. The outer peaks shift to
away from zero energy, while the inner ones merge and form a small ZBCP.

Including a barrier in the SW, both CPW and HPW, $\theta_M = 0$ cases [Figs.~\ref{fig:magfield_inclbar}(a)-\ref{fig:magfield_inclbar}(c)] 
does not change the
behaviour qualitatively, but the overall structure is more pronounced.
In the HPW, $\theta_M=\pi/2$ case [Fig.~\ref{fig:magfield_inclbar}(d)],
however, the spectrum changes from a
plateau to a three-peak structure.
The CPW, $\theta_M=0$ and HPW, $\theta_M=\pi/2$ cases can be distinguished
by looking at the relative peak height of the ZBCP.

\section{Summary}
We have investigated the conductance of a N/F/S junction 
with various pair potentials as a function of ferromagnetic properties
(thickness, magnetization strength and direction).
The SW and CPW, $\theta_M=0$ cases are similar, although the latter
shows a more rounded conductance due to the angle dependence of the pair
potential. 
We found that the cases where the $d$-vector is perpendicular to the
 magnetization direction (CPW, $\theta_M=\pi/2$ and HPW, $\theta_M=0$)
 are identical. In these cases, the opposite spins parts of the Hamiltonian
 are decoupled and therefore,
 they are insensitive to the ferromagnet thickness and magnetization
 strength.
The cases where the $d$-vector is parallel to the magnetization
direction are very different due to a more complex structure.
The main difference is that CPW, $\theta_M=0$ converges to a zero
energy peak for $k_FL=15$, while HPW, $\theta_M=\pi/2$ shows a dip.
In the presence of an external magnetic field,
	the evolution of the conductance spectra depends on the
	pairing symmetry.
	In particular, the CPW, $\theta_M=0$ case gives an accidental ZBCP.
	The central dip in the CPW, $\theta_M=\pi/2$ and HPW, $\theta_M=0$ cases
	remains. In the HPW, $\theta_M=\pi/2$ case, the structure depends on the
	barrier strength; a plateau or three peaks.

For future research, it would be interesting to take higher applied
magnetic fields into account by including Abrikosov vortices.
To obtain a more accurate representation of the Sr$_2$RuO$_4$,
tunneling spectroscopy can be simulated using a multiband
model.\cite{Yada,Kawai}

\acknowledgements
This work  was supported by JSPS Core-to-Core program ``Oxide Superspin" international network, a Grant-in-Aid for Scientific Research on Innovative Areas Topological Material Science JPSJ KAKENHI (Grants No. JP15H05851,No. 15H05852, No. JP15H05853, and No. JP15K21717), and a Grant-in-Aid for Scientific Research B (Grant No. JP15H03686 and JP18H01176). It  was also supported by JSPS-RFBR Bilateral Joint Research Projects and Seminars Grant No. 17-52-50080, the Ministry of Education and Science of the Russian Federation Grant No. 14.Y26.31.0007 and by joint Russian-Greek projects RFMEFI61717X0001 and T4$\Delta$P$\Omega$-00031.

\appendix
\section{Doppler shift \label{app:doppler} }
In the presence of a magnetic field $\bs{H} = \bs{\nabla}\times \bs{A}$
the canonical momentum operator $\bs{p}$ is replaced by the kinetic
momentum operator 
$\bs{\pi}=\bs{p} - e\bs{A}(\vec{r})/c$.
As a result, the quasi particle kinetic energy $\xi_k$ becomes
\begin{align}
\xi_k = \frac{1}{2m} \bs{\pi} \cdot \bs{\pi} -\mu
= - \frac{\hbar^2}{2m} \left( \bs{\nabla}
- i \frac{|e|}{\hbar c}\bs{A}\right)^2 -\mu.
\end{align}
where $\mu$ is the chemical potential.
In the weak-coupling limit ($\Delta_0 \ll \mu$), this can be approximated by
\begin{align}
\xi_k \approx - \frac{\hbar^2\nabla^2}{2m}
- i \frac{\hbar|e|}{mc}\bs{\nabla}\cdot\bs{A} - \mu.
\label{weakcoupl}
\end{align}
In our case, an external magnetic field $\bs{H}$ is applied in the
$x$-direction. Hence, the magnetic field and vector potential for $z\ge 0$
are approximately\cite{fogelstrom}
\begin{align}
&\bs{H}(z) = H e^{-z/\lambda_L} \boldsymbol{e}_x, \\
&\bs{A}(z) = -H\lambda_L e^{-z/\lambda_L} \boldsymbol{e}_y,
\end{align}
where $\lambda_L$ is the London penetration depth.
The spatial dependence of $\bs{A}$ is characterized by $\lambda_L$,
whereas the Cooper pair wave function is characterized by the
coherence length $\xi_0$.
In the type-II limit ($\lambda_L/\xi_0 \gg 1$), the spatial
dependence of $\bs{A}$ does not change the differential
conductance. Therefore we introduce the constant
vector potential\cite{tanakajpsj}
\begin{align}
	\bs{A}(z) \approx -H \lambda_L \boldsymbol{e}_y.
\end{align}
This linear response is only valid in the absence
	of vortices, i.e. for small magnetic fields ($H\le 0.6H_{c}$).
Assuming plane waves in the $x$- and $y$-direction, the wave function
can be written as $\psi(x,y,z) = \psi(z)e^{ik_x x}e^{ik_y y}$,
such that Eq.~(\ref{weakcoupl}) becomes
\begin{align}
\xi_k = - \frac{\hbar^2}{2m}\frac{\partial^2}{\partial z^2}
+ \frac{\hbar^2 k_\parallel^2}{2m}
- \frac{\hbar|e|}{mc}H \lambda_L k_y -\mu,
\label{eq:kinenergy}
\end{align}
where $k_\parallel^2 = k_x^2+k_y^2$.
Defining $\mu' \equiv \mu - \hbar^2k_\parallel^2/2m$
and substituting $H_c = \phi_0/\pi\xi_0 \lambda_L$,
$\phi_0 = \pi \hbar c/|e|$, $\xi_0 = \hbar v_F/\Delta_0$,
$k_y=k_\parallel \sin\varphi$ and $v_F = \hbar k_F/m$,
Eq.~(\ref{eq:kinenergy}) can be written as
\begin{align}
\xi_k = - \frac{\hbar^2}{2m}\frac{\partial^2}{\partial z^2}
- \mu'
- \Delta_0\frac{H}{H_c} \frac{k_\parallel}{k_F}\sin\varphi,
\label{eqdoppler}
\end{align}
where $H_c$ is the thermodynamical critical field.

\section{Numerical method}
Substituting wave functions Eqs.~(\ref{eq:waveN}) and
(\ref{eq:waveF}) into boundary condition Eq.~(\ref{eq:bc1})
gives
\begin{equation}
\vec{i} + \vec{r} = \check{A}\left( \vec{f}_p + \vec{f}_n \right). 
\label{A:bc1}
\end{equation}
We do the same with boundary condition Eq.~(\ref{eq:bc2})
and divide by $ik_0$ for normalization, where we define
$k_0$ as the momentum in the normal metal, i.e.
$k_0 = \sqrt{2m_N\mu}/\hbar$.
The second boundary condition becomes
\begin{equation}
\left( \frac{k_N}{k_0}\check{\tau}_0
-2i\check{Z}_1 \right)\vec{i}
- \left( \frac{k_N}{k_0}\check{\tau}_z 
+2i\check{Z}_1 \right)\vec{r}
= \check{A}\check{Q} \left( \vec{f}_p - \vec{f}_n \right)
\label{A:bc2} 
\end{equation}
where $\check{Q} = \mathrm{diag}[~
k_F^+, \, ~
k_F^-, \, ~
k_F^+, \, ~
k_F^-~]/k_0$ and $\check{Z}_1$ is the dimensionless barrier strength
of the first interface, given by
\begin{equation}
\check{Z}_1 = \frac{m(z) \check{F}_1 }{\hbar^2 k_0}.
\end{equation}
We substitute wave functions Eqs.~(\ref{eq:waveF}) and (\ref{eq:waveS})
into the third boundary condition, Eq.~(\ref{eq:bc3}), to obtain
\begin{equation}
\check{A}\check{K}_F^{L+} \vec{f}_p + \check{A}\check{K}^{L-}_F \vec{f}_n 
= \check{U}\check{K}_S^L \vec{t},
\label{A:bc3}
\end{equation}
where we used  $\check{K}_F^{L\pm} = \left.\check{K}^\pm_F\right|_{z=L}$
and 
$\check{K}_S^{L\pm} = \left.\check{K}^\pm_S\right|_{z=L}$ for abbreviation.
Similarly, from Eq.~(\ref{eq:bc4}), we get
\begin{align}
&\check{A}\check{Q} \left( \check{K}_F^{L+} \vec{f}_p
- \check{K}^{L-}_F \vec{f}_n \right)
- 2i\check{Z}_{\mathrm{SO}} \check{A}\left( \check{K}_F^{L+} \vec{f}_p
+ \check{K}^{L-}_F \vec{f}_n \right) \nonumber \\
& \hspace{5cm} = \frac{k_S}{k_0}
\check{U} \check{\tau}_z \check{K}^L_S  \vec{t},
\label{A:bc4}
\end{align}
where $\check{Z}_{\mathrm{SO}}$ is the dimensionless spin-orbit coupling
strength at the second interface, defined as
\begin{equation}
\check{Z}_{\mathrm{SO}} = \frac{m(z) \check{F}_{\mathrm{SO}} }{\hbar^2 k_0}.
\end{equation}
Eqs.~(\ref{A:bc1}), (\ref{A:bc2}), (\ref{A:bc3}), (\ref{A:bc4})
form a system of 16 equations with 16 unknowns.
Substituting Eqs.~(\ref{A:bc3}) and (\ref{A:bc4})
into one another, we can write $\check{M}_1 \vec{f}_p =  \check{M}_2 \vec{f}_n$, with
\begin{align*}
& \check{M}_1 = \check{A} \check{Q} \check{K}_F^{L+}
- 2i\check{Z}_{\mathrm{SO}}\check{A}\check{K}_F^{L+}
- \frac{k_S}{k_0} \check{U} \check{\tau}_z
\check{U}^{-1} \check{A} \check{K}_F^{L+}, \\
& \check{M}_2 = \check{A} \check{Q} \check{K}_F^{L-}
+ 2i\check{Z}_{\mathrm{SO}}\check{A}\check{K}_F^{L-}
+ \frac{k_S}{k_0}\check{U}\check{\tau}_z
\check{U}^{-1} \check{A} \check{K}_F^{L-}.
\end{align*}
Combining this with Eq.~(\ref{A:bc2}), we can express
$\vec{f}_p$ and $\vec{f}_n$ in terms of $\vec{i}$ and $\vec{r}$ as
\begin{align}
\vec{f}_n &= \check{M}_3^{-1} (\vec{i} + \vec{r}) \label{eq:fn} \\
\vec{f}_p &= \check{M}_1^{-1} \check{M}_2
\check{M}_3^{-1} (\vec{i} + \vec{r}) \label{eq:fp},
\end{align}
where $\check{M}_3 = \check{A}\left( \check{M}_1^{-1} \check{M}_2
+ \check{\tau}_0 \right)$.
Substituting Eqs.~(\ref{eq:fn}) and (\ref{eq:fp}) into
Eq.~(\ref{A:bc1}), we find that
\begin{equation}
\vec{r} =  \check{M}_4^{-1}\check{M}_5 \vec{i}
\end{equation}
with
\begin{align*}
\check{M}_4 &= \frac{k_N}{k_0}\check{\tau}_z +2i\check{Z}_1
- \check{A}\check{Q} \left( \check{M}_1^{-1} \check{M}_2
- \vec{\sigma}_0 \right) \check{M}_3^{-1}, \\
\check{M}_5 &= \frac{k_N}{k_0}\check{\tau}_0
-2i\check{Z}_1 + \check{A}\check{Q} \left( \check{M}_1^{-1} \check{M}_2
- \vec{\sigma}_0 \right) \check{M}_3^{-1}.
\end{align*}
Using the $\vec{r}$ coefficients, the conductance can be determined
by Eq.~(\ref{eq:cond}).

\section{Rotation of spin quantisation axis \label{appendixrotate}}

To discuss the spin of Cooper pairs, it is convenient to rotate the spin
quantization axis such that the new $z$-axis is parallel to the magnetization $\bs{M}$.
In our case, $\bs{M}$ is in the $xz$-plane in spin space.
Therefore, the rotation should be around the $y$-axis in spin space,
which is carried out by the unitary operator
\begin{align}
  \hat{U}(\theta_M) 
  &= \exp \left[ i ({\theta_M}/{2}) \hat{\sigma}_y \right] \\[1mm]
  &= \hat{\sigma}_0 \cos \left( {\theta_M}/{2} \right) 
  +i \hat{\sigma}_y \sin \left( {\theta_M}/{2} \right), 
  \label{eq:spin:1}
\end{align}
with which we can rotate spin space by an angle $\theta_M$. 
The unitary matrix in Eq.~(\ref{eq:spin:1}) 
satisfies $\hat{U}^* = \hat{U}$ and therefore,
the unitary matrix in Nambu space is given by $\check{U} =
\mathrm{diag}[ \hat{U}, \hat{U}^* ] =
\mathrm{diag}[ \hat{U}, \hat{U} ]$. 
The BdG equation changes accordingly and becomes
\begin{align}
  \check{H} \Psi
  =       {E} \Psi 
  \hspace{4mm}
  \rightarrow
  \hspace{4mm}
  &\check{H}' \tilde{\Psi}
  =       {E} \tilde{\Psi}, 
\end{align}
with 
\begin{align}
  & \tilde{\Psi} = \check{U}{\Psi}, \hspace{4mm}
  \check{H}' = \check{U}\check{H}\check{U}^\dagger, 
	\\ 
	& \Psi = 
  \Big[~
    {\psi}_{\uparrow}         ~ ~
    {\psi}_{\downarrow}         ~ ~
    {\psi}_{\uparrow}^\dagger ~ ~
    {\psi}_{\downarrow}^\dagger ~ 
  \Big]^T.
\end{align}
\begin{figure}[tb]
	\centering
	\includegraphics[width=0.9\linewidth]{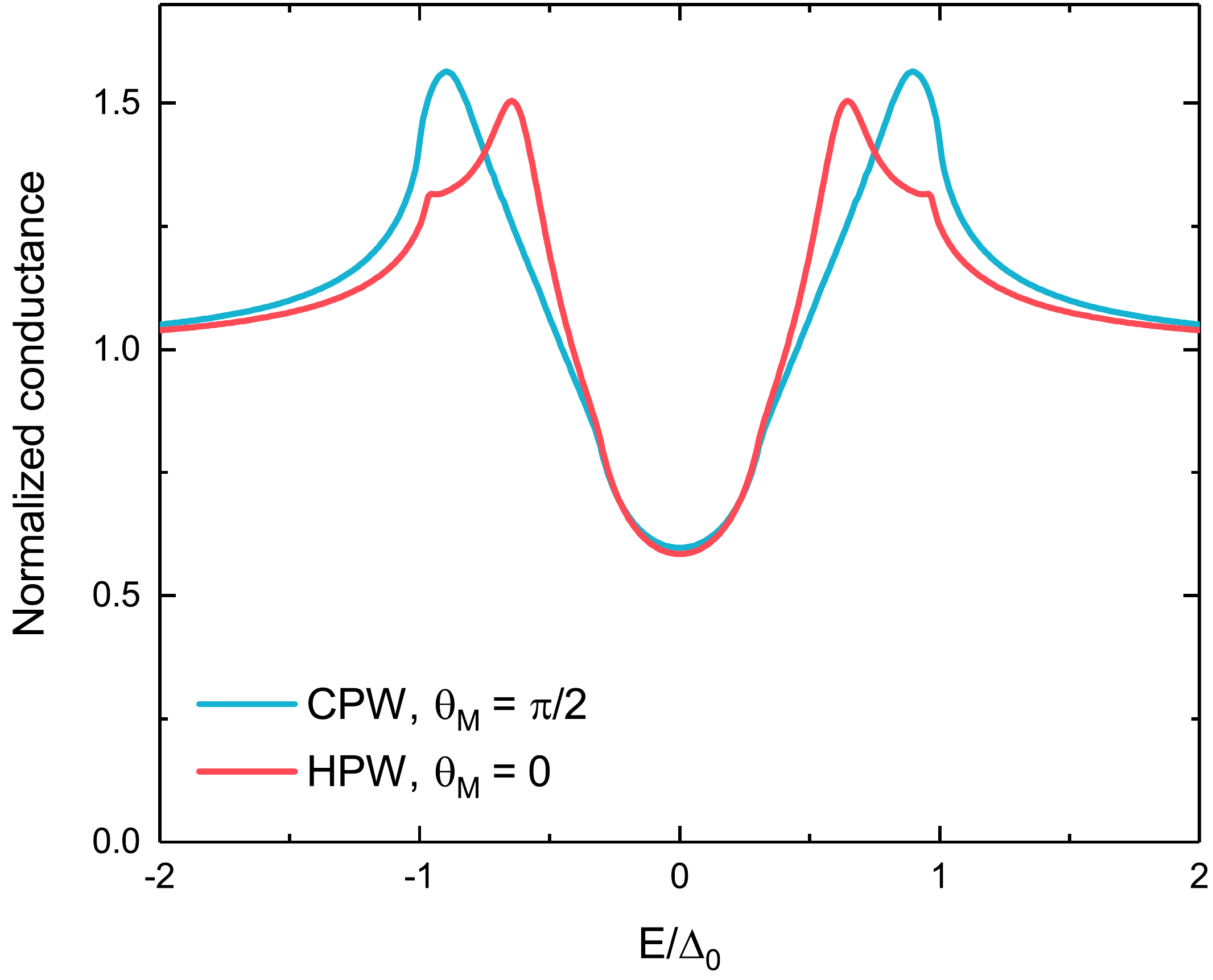}
	\caption{The dimensionless tunneling conductance 
		for CPW, $\theta_M=\pi/2$ and HPW, $\theta_M=0$,
		including spin-orbit coupling $Z_{SO}=1$.
		$Z_1 = 0.8$, $X = 0.6$, $k_FL = 11$.
	}
	\label{fig:SOC}
\end{figure} 

Only the magnetization term depends on spin in the single-particle
 Hamiltonian
$\hat{h}(z)$.  In the new spin basis, the magnetization term
for particles and holes is given by, respectively,
\begin{align}
   &\hat{U} 
  \big( \bs{M} \cdot \hat{\bs{\sigma}} 
  \big) \hat{U}^\dagger = M \hat{\sigma}_z, \\
   & \hat{U} 
  \big( -\bs{M} \cdot \hat{\bs{\sigma}}^* 
  \big) \hat{U}^\dagger = -M \hat{\sigma}_z.
\end{align}
The pair potential in the new spin space is 
\begin{align}
\left[
\begin{array}{cc}
&
\underline {\hat{\Delta}}  _{ \bs{k}_\parallel} \\[2mm]
-\underline {\hat{\Delta}}^*_{-\bs{k}_\parallel} &
\\
\end{array}\right]
\hspace{2mm}
\rightarrow
\hspace{2mm}
\left[
\begin{array}{cc}
&
\hat{U}
\underline {\hat{\Delta}}  _{ \bs{k}_\parallel}
\hat{U}^\dagger \\[2mm]
-\left[
\hat{U}
\underline {\hat{\Delta}}  _{-\bs{k}_\parallel}
\hat{U}^\dagger \right]^* &
\\
\end{array}\right], 
\end{align}
where we used the relation $\hat{U}^* = \hat{U}$. 
The superconducting pair potential $\underline{\hat{\Delta}}  _{ \bs{k}_\parallel}$ is transformed to
\begin{align}
&\hat{U} \underline{\hat{\Delta}}  _{ \bs{k}_\parallel}
\hat{U}^\dagger  \nonumber\\
&=\left\{ \begin{array}{ll}
\Delta_0 i \hat{\sigma}_y & \text{for SW, } \\[2mm]
\Delta_0
\big( \bar{k}_x + i \chi \bar{k}_y \big)
\big[ \cos \theta \hat{\sigma}_x 
+ \sin \theta \hat{\sigma}_z \big]
& \text{for CPW, } \\[2mm]
\Delta_0
\big( \bar{k}_x \hat{\sigma}_0 
+ i \bar{k}_y 
[ \cos \theta \hat{\sigma}_z 
- \sin \theta \hat{\sigma}_x ] \big)
& \text{for HPW. } \\[2mm]
\end{array} \right.
\end{align} 
If we substitute $\theta_M = 0, \pi/2$, these expressions
reduce to the pair potentials in table~\ref{tablepairpotential}
in the main text.

We focus on CPW, $\theta_M = \pi/2$ and HPW, $\theta_M = 0$.
In both cases, the magnetization $\bs{M}$ is perpendicular to the
$d$-vector. In other words, $\bs{M}$ and the total spin of
Cooper pairs are collinear. 
Therefore, the magnetization does not destroy the Cooper pairs.
The $4\times 4$ Hamiltonian matrix can be reduced to two $2 \times 2$ matrices: 
\begin{align}
  \mathcal{H} 
  &= \frac{1}{2} \, \sum_{\bs{k}_\parallel} \int 
  \tilde{\Psi}^\dagger (z)
  \check{H}_{B}(z) 
  \tilde{\Psi}         (z)
  \, dz, \nonumber \\[0mm]
  &= \frac{1}{2} \, \sum_{\bs {k}_\parallel} \int \sum_{\alpha = \pm 1}
  \tilde{\Psi}^\dagger_\alpha
  \left[ \begin{array}{cccc}
    \xi + \alpha M & \Delta_{\alpha, k_\parallel} \\[2mm]
    -\Delta_{\alpha,-k_\parallel}^* & -(\xi + \alpha M) \\[1mm]
  \end{array}\right]
  \tilde{\Psi}_\alpha 
  \, dz. 
  \label{eq:22}
\end{align}
We have introduced a new basis which depends on the spin sector $\alpha$: 
$
\tilde{\Psi}_\alpha(z) = 
[\,
\tilde{\psi}_{\alpha}         (z) \hspace{2mm} 
\tilde{\psi}_{\alpha}^\dagger (z)              \,]. 
$ 
Eq.~(\ref{eq:22}) implies that the system can be decomposed into the
spin-up ($\alpha=1$) and spin-down ($\alpha=-1$) subsystems, where we have
redefined the up and down spins for the new spin quantization axis. 
 The $\alpha$-dependent pair potential is given by 
\begin{align}
\Delta_{\alpha, k_\parallel}
= \left\{ \begin{array}{ll}
\alpha{\Delta_0} e^{i        \phi} & \text{for CPW, }\theta_M=\pi/2, \\[2mm]
{\Delta_0} e^{i \alpha \phi} & \text{for HPW, }\theta_M = 0, \\[2mm]
\end{array} \right.
\end{align}
where we fix $\chi = 1$. In the CPW, $\theta_M=\pi/2$ case,
the chiralities for up- and down-spin sectors are the same,
while the signs of the $\alpha$-dependent pair potential are opposite.
In the HPW, $\theta_M=0$ case, the chiralities are opposite,
while the signs of the $\alpha$-dependent pair potential are equal.
Therefore, as far as there is no perturbation which mixes the spins or
depends on the chirality ($e.g.$, spin-active interface, spin-orbit coupling and perturbations which breaks translational symmetry
in the $x$ and/or $y$ direction such as walls and impurities), it is impossible to distinguish these two cases.

This is demonstrated in Fig.~\ref{fig:SOC}, where we introduced
spin-orbit coupling at the F/S interface by setting $Z_{SO}=1$.
In the absence spin-orbit coupling, these two graphs overlapped
as seen in Figs.~\ref{fig:magnetization}(c), \ref{fig:length}(c),
\ref{fig:magfield_nobar}(c), and \ref{fig:magfield_inclbar}(c). However, in
Fig.~\ref{fig:SOC}, 
we can see that they are indeed slightly different.

\bibliography{ref1} 

\begin{thebibliography}{78}
\bibitem [{not()}]{notesum}%
  \BibitemOpen
  \href@noop {} {}\bibinfo {note} {We note that Eq.~(\ref{sumsigma}) is still
  valid in the presence of spin-rotation. SOC only influences the transport
  coefficients and does not affect the validity of
  Eq.~(\ref{sumsigma}).}\BibitemShut {Stop}%
\bibitem [{int()}]{interference}%
  \BibitemOpen
  \href@noop {} {}\bibinfo {note} {There are two types of the zero-energy
  conductance peaks. One is a conductance peak which appears accidentally at
  zero-energy, the other is a conductance peak which should appear at the
  zero-energy because of, for example, the Andreev interface/surface bound
  state originating from the topology of the bulk wave functions. A zero-energy
  peak is not expected on the surface perpendicular to the $c$ direction of
  Sr$_2$RuO$_4$. Moreover, if the peak in Fig.~3a is caused by, for example,
  the Andreev bound states in F, the energy of the conductance peek is not
  affected by the details of a system such as the thickness. In Fig.~3a, the
  peeks at $\pm \Delta_0$ for $L=0$ move in the $\mp E$ direction with
  increasing $L$. At $k_F L = 15$, these two peaks merge. For a longer $L$, the
  two peaks just pass each other.}\BibitemShut {Stop}%
\end{thebibliography}

\end{document}